\documentclass[11pt]{article}
\usepackage{graphicx}

\begin{document}

\title{TeVeS/MOND is in harmony with gravitational redshifts in galaxy clusters}
\author{J. D. Bekenstein\footnote{bekenste@vms.huji.ac.il} and R.H. Sanders\footnote{sanders@astro.rug.nl}}
\date{\today}
\maketitle
\begin{abstract}
Wojtak, Hansen and Hjorth have recently claimed to confirm general relativity and to rule out the tensor-vector-scalar (T$e$V$e$S) gravitational theory based on an analysis of the gravitational redshifts of galaxies in 7800 clusters.  But their ubiquitous modeling of the sources of cluster gravitational fields with Navarro-Frenk-White  mass profiles is neither empirically justified out to the necessary radii in clusters, nor germane in the case of T$e$V$e$S.  Using MONDian isothermal sphere models consistently constructed within MOND (equivalent to T$e$V$e$S models), we can fit  the determined redshifts no worse than does general relativity with dark halos.   Wojtak, Hansen and Hjorth's work is further marred by confusion between the primitive mu-function of T$e$V$e$S and the MOND interpolation function.
\end{abstract}
In a recent article Wojtak, Hansen and Hjorth$^1$ (WHH) analyze the pattern of spectroscopic redshifts for galaxies in 7800 clusters from the 7th data release of the Sloan Digital Survey.  Motion of a galaxy in a cluster engenders a red- or blueshift of its spectral lines; the equivalent velocities are typically on the order of 600\,km\,s$
^{-1}$. WHH make a good case that they are able to dig out the superimposed gravitational redshift, which is equivalent to a velocity on the order 10\,km\,s$^{-1}$,  by stacking data from all clusters and measuring the displacement of the redshift distribution's centroid with growing radial coordinate $r$ in the typical cluster.

However useful their technique may turn out to be, WHH draw from it erroneous conclusions: that general relativity (GR) is confirmed by the observed redshifts pattern, that the $f(R)$ gravity theory$^{2}$ may well be compatible with the data, and that the Tensor-Vector-Scalar theory$^3$ (T$e$V$e$S), an early  relativistic gravity implementation of the MOND paradigm$^4$ (which claims to describe galaxy dynamics without dark matter), is ruled out.  

Let us recall that in the mentioned theories, as well as in any other metric gravitational theory endowed with the usual weak field limit, $g_{00}\approx -c^2-2\Phi$, the gravitational redshift is quantified, to first order, by the same gravitational potential $\Phi$ which controls motion of galaxies in their parent clusters, etc.$^5$  The gravitational redshift suffered by a galaxy situated at radial position $r$ should follow from the formula $z=-\Phi(r) c^{-2}$ [with $\Phi(\infty)=0$; gravitational potential is negative].  This is a straight consequence of the Einstein equivalence principle which is implemented in all such theories.$^6$ 

WHH determine $\Phi(r)$ in a cluster, by matching the dispersion of galaxy velocities in their stacked cluster with the best fitting NFW halo model (which emerges from cosmic N-body simulations within GR$^7$), and appealing to an assumption about the predominant shapes of galaxy orbits.  Actually they determine the gravitational shift relative to that of light emerging from the center of the cluster, so their result is presented as a blueshift: $\Delta z = [\Phi(0)-\Phi(r)]/c^2 <0$.  Now were $\Phi(r)$ inferred directly from galaxy velocity dispersions, the curve $\Phi(r)$ vs $r$ would have to be the same for all metric theories with a standard weak field limit.  So how did WHH get their Fig.~2 where the redshift curve for T$e$V$e$S lies well below GR's?

While the gravitational redshift is in no sense a test of competing theories of gravity (as long as all
respect the equivalence principle), it can be a test of the mass distribution in the outer parts of clusters, i.e.,  the mass distribution that may be implied by a particular theory.  For clusters in general, the mass distribution can be measured dynamically, by the motion of galaxies or the density and temperature distribution of hot gas, but these techniques work out to $r=$1--2 Mpc at most.  Yet WHH purport to observationally determine the gravitational redshift, and hence the potential, out to 6 Mpc.   What do they assume about the mass distributions in these outer regions in light of competing theories of gravity?

For GR with dark matter  WHH {\it assume} that the mass distribution out to 6 Mpc, four times the typical virial radius, is well described by the popular NFW profile.$^7$   They pick that NFW model, specified by a virial radius and a concentration parameter, which best fits the velocity dispersion data of the average cluster within about one Mpc of the center.  Now in a NFW halo beyond the virial radius of 1--2 Mpc, the density mass distribution falls as $1/r^3$.   It must be emphasized that this behavior is an assumption; there is no independent tracer of the mass distribution out to 6 Mpc where the cluster may not be in virial equilibrium and may still be accreting.  

MOND/T$e$V$e$S is less flexible;  as presently understood the theory  unavoidably yields an asymptotically logarithmically rising  potential which is equivalent to a ``phantom" dark matter density falling as $1/r^2$ in the outer
regions (phantom in the sense of dark matter that one would suppose to be present were one to analyze the problem Newtonianly). Thus the \emph{effective} density declines more slowly than for the assumed NFW profile. One might suppose that this is the reason for the different curves in Fig.~2 of WHH: MOND/T$e$V$e$S produces more potential in the outer regions of clusters than does GR for an equal true mass distribution in the inner regions. In their supplementary material, WHH admit that the
mass distribution is uncertain in the outer regions but claim that the
GR result is fairly insensitive to the actual exponent of
a power law density distribution.

But the actual problem is primarily one of scaling. In their test of the MOND/T$e$V$e$S scenario WHH again assume a NFW mass distribution, but one
scaled down to 80\% of the Newtonian dynamical mass.  That is to say, the discrepancy between the observed and dynamical masses is reduced from a factor of six to a factor of five (in clusters some unseen matter is also required by MOND$^8$).  This assumption is  based on an analysis of clusters 
by Pointecouteau and Silk.$^9$  However, as pointed out by Milgrom,$^{10}$ these authors used too small a value of $a_0$.  The actual ratio of MOND to Newtonian mass should be more like 50\%, and this factor decreases with radius.$^{11}$

However, the NFW profile is not particularly germane to the MOND/T$e$V$e$S world view; the profile arises in GR cosmological simulations.  Therefore, we have calculated $\Delta z$  using instead as source the mass distribution of a MONDian  isothermal sphere with a central velocity dispersion of 600\,km\,s$
^{-1}$.  The MOND potential difference was calculated using the
MOND equation$^4$ (here and henceforth $\Phi'\equiv \partial\Phi/\partial r$ and $\Phi_N$ is the usual Newtonian potential)
\begin{equation}
\tilde\mu(\Phi')\,\Phi'= \Phi_N{}',
\end{equation}
with the ``standard'' interpolating function$^{4,12}$
\begin{equation}
\tilde\mu(\Phi')=\frac{\Phi'/a_0}{\sqrt{1+\Phi'^2/a_0^2}}\ .
\end{equation}
 The total mass of our MONDian isothermal sphere is $1.45\times 10^{14}$ M$_\odot$ (MONDian isothermal isothermal
spheres have finite mass).  By comparison, we require a mass of about $4\times 10^{14}$ M$_\odot$ for the NFW halo out to 6 Mpc to emulate WHH's GR results.
Thus the MOND dynamical mass is only 36\% of the Newtonian dynamical mass
within 6 Mpc, not 80\%.   

The results are displayed in our Fig.~1 along with the data of WHH
(comparable to their Fig.~2).  The solid curve is the MOND prediction and the dashed line is that of GR plus NFW halos.  We see that, while the MOND curve does lie below the GR curve, both are consistent with the observations.   The data does not distinguish between the predictions of the two theories because the natural  mass distribution in the outer regions is different for GR and MOND.
The two curves almost coincide within 1 Mpc from the center, as they should,  since the mass distributions and potentials coincide here becoming different only in the outer regions.

\begin{figure}[]
\centering
\includegraphics[width=0.6\textwidth]{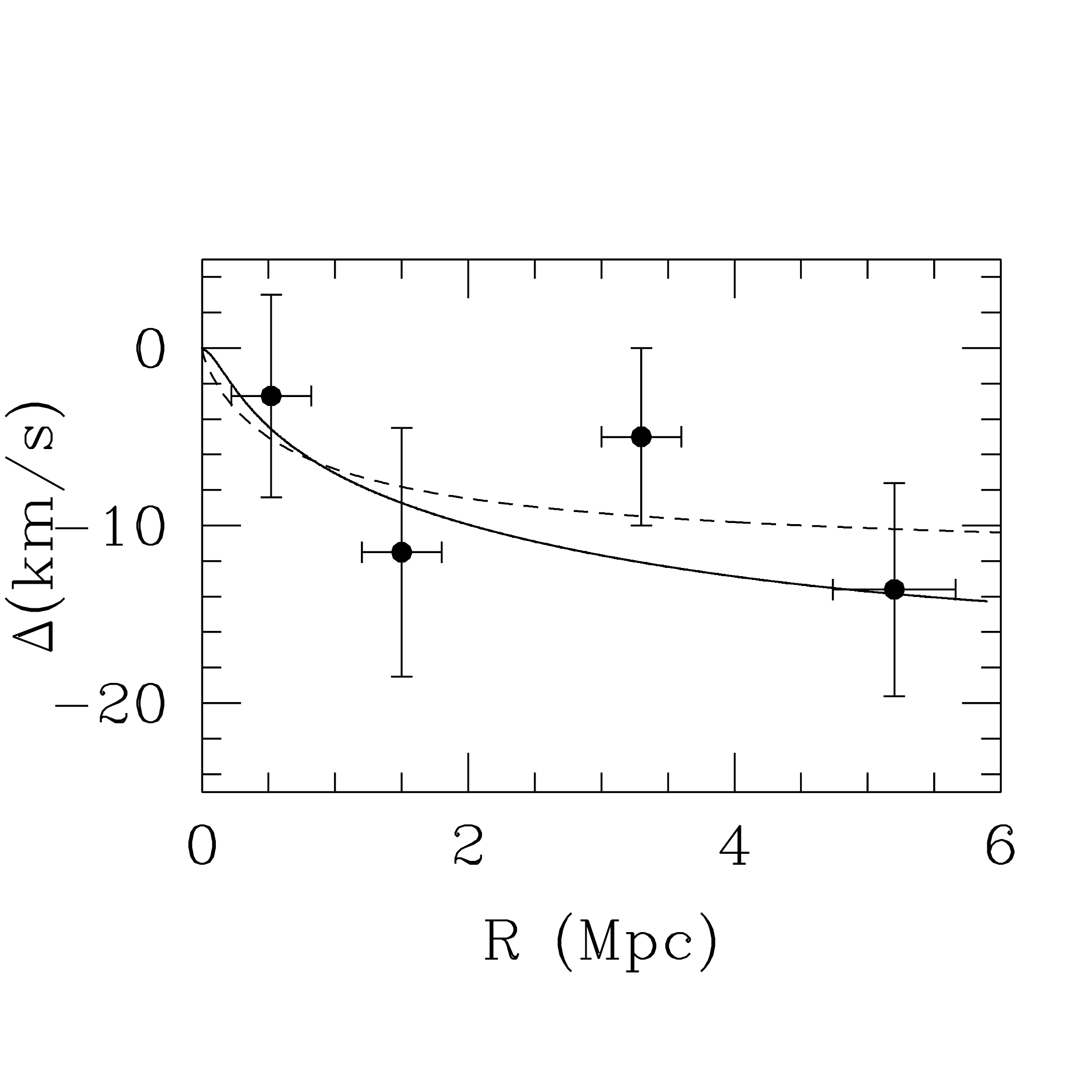}
\caption{Gravitational redshift with respect to the center of the cluster
following WHH.  The dashed line is that of GR plus the NFW model
which best fits the run of velocity dispersion in the central regions of
the average cluster.  The solid line is that of the MOND/TeVeS isothermal sphere with a central velocity dispersion of 600\,km\,s$
^{-1}$.   The data points are from WHH's paper.} 
 \end{figure}

Independent of the issue of the mass distribution in the outer reaches of a cluster, WHH made a conceptual error in their  specifically T$e$V$e$S calculation of the gravitational redshift. In GR $\Phi$ coincides with $\Phi_N$, the latter sourced by baryonic and dark matter.   Although T$e$V$e$S comprises two metrics, primitive and physical, light and matter propagate exclusivelys on the physical metric.  The potential $\Phi$ is related to the physical metric in just the way  $\Phi_N$  is related to GR's metric. T$e$V$e$S also includes a scalar field $\phi$ satisfying a nonlinear field equation, formulated on the primitive metric, which is defined with help of a specific function $\mu$ of the gradient of $\phi$.  
This equation has the consequence that in a spherical cluster 
\begin{equation}
\mu(\phi')\,\phi'=\frac{k}{4\pi}\, \Phi_N{}',
\end{equation}
where $k$ is a small coupling constant$^{3}$ (WHH assume $k=0.01$).  As in the MOND paradigm $\Phi_N$ here is considered to be sourced only by the baryonic matter.  The two metrics in T$e$V$e$S are so related that to linear order
\begin{equation}
\Phi=\Phi_N+\phi.
\end{equation}

By contrast to WHH's procedure to be summarized below, in our T$e$V$e$S analysis of the gravitational redshift we have worked directly with the full MOND equation, Eq.~(1), using the ``standard" form of $\tilde{\mu}$.  This procedure is entirely permissible when spherical symmetry obtains since, as shown by Eq.~(4),  the scalar $\phi$ plus the Newtonian potential combine in T$e$V$e$S to provide the effective weak field potential $ \Phi$, which yields directly the gravitational redshift.  It is unnecessary to consider the scalar field $\phi$ separately as WHH did.

While WHH took $\Phi_N$ in their T$e$V$e$S analysis to be a fraction 80\% of the $\Phi_N$ required by the velocity dispersions in Newtonian gravity, we have argued above that the fraction should be more like a third.   To determine $\Phi$ from Eq.~(4) WHH computed $\phi$ through Eq.~(3) using the prescription
\begin{equation}
\mu(\phi')=\frac{\surd k\,l\phi'}{1+\surd k\,l\phi'}\ ,
\end{equation} 
where $l$ is the T$e$V$e$S scale of length (whose relation to the MOND acceleration scale WHH take to be $a_0=\surd\kappa/(4\pi l)$ based on Ref.~3).  They justify this choice of $\mu$ because in MOND's versatile equation~(1)
an oft used choice is the ``simple'' interpolating function$^{12}$ 
\begin{equation}
\tilde\mu(\Phi')=\frac{\Phi'/a_0}{1+\Phi'/a_0}\ .
\end{equation}

\begin{figure}[]
\centering
\includegraphics[width=0.6\textwidth]{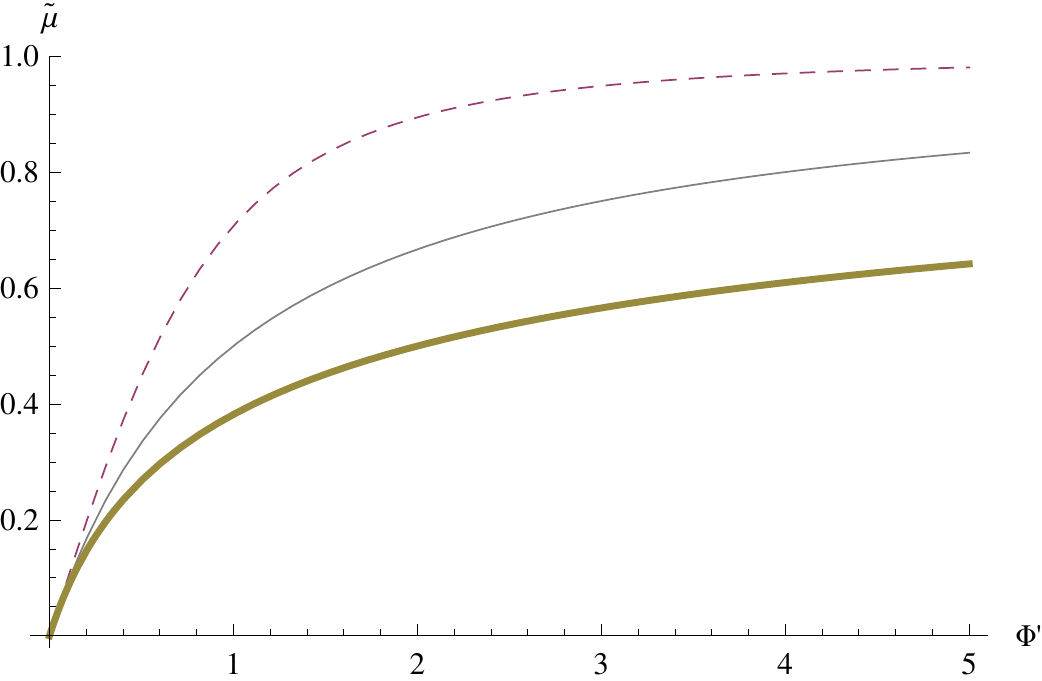}
\caption{In this plot of $\tilde\mu$ vs $\Phi'$ (in units of $a_0$) the thick solid curve depicts the interpolation function for WHH's choice of $\mu(\phi')$ [Eq.~(5)] and $k=0.01$, while the light solid and dashed curves are for the well tested ``simple" and ``standard" interpolation functions, Eqs.~(6) and (2), respectively.}
\end{figure}

But there is a confusion here. In T$e$V$e$S $\mu$ and $\tilde\mu$ are different functions.$^3$   WHH's choice (5) corresponds, by Eqs.~(59)-(61) of Ref.~3, to the relation
\begin{equation} 
\Phi'/a_0=\frac{\tilde\mu}{1-\tilde\mu}\ \frac{1}{1-(1+\frac{k}{4\pi})\tilde\mu}\ .
\end{equation}
This being quadratic, WHH's choice of $\mu(\phi')$ corresponds to a \emph{double valued} interpolation function $\tilde\mu(\Phi')$.  One of its branches is physically ruled out because it gives $\tilde\mu>1$. The permissible branch is plotted in Fig.~2 for WHH's choice $k=0.01$.  For a given source it implies a significantly stronger gravitational field than do either of the well tested ``simple'' and ``standard'' functions also shown [see Eq.~(1)].  It is thus no wonder that WHH predict too deep a cluster potential, as reflected by their T$e$V$e$S gravitational redshift curve in Fig.~2 of WHH. 

But how does WHH's implied MOND interpolation function, when combined with a mass distribution natural to MOND, fare in light of the cluster data?  We have repeated our calculation using a MONDian isothermal sphere source with MOND's equation (1), but this time with the interpolation function represented by the solid thick curve in our Fig.~2.   A source with 80\% of the Newtonian mass does lead to agreement with WHH's gravitational redshift \emph{prediction} of -22\,km\,s$ ^{-1}$ at the cluster center, but pushes the predicted central galaxy velocity dispersion up to 850\,km\,s$ ^{-1}$, a number excluded by the data.  The dispersion can be brought down to the observed level by reducing the source mass to 20\% of Newtonian, in which case the central gravitational redshift is down to -13.6\,km\,s$ ^{-1}$ and the redshift curve is a bit closer to the GR one than shown in Fig.~1.  Use of WHH's implied interpolation function undeniably eases MOND'd need for dark matter in clusters.  It should be remarked, however, that WHH's interpolation function has hardly been tested with galaxy rotation curves, an areas in which MOND with the usual interpolation functions is very successful.

To conclude, WHH's far reaching claims do not stand up under close scrutiny.  The NFW profile which they assume everywhere is not supported by independent evidence out to the radial distances which they explore.  And its use in the context of T$e$V$e$S is unnatural.  In our test  of T$e$V$e$S/MOND a mass distribution consistently constructed within MOND gives predictions for the redshift in no way inferior to those obtained by WHH for GR.  So WHH's claim that GR explains the gravitational redshift measurements better than does T$e$V$e$S/MOND is baseless. Nonetheless, increased precision in the measurement of the gravitational redshift might make it possible in the future to distinguish between GR with NFW halos and TeVeS/MOND.  

We thank M. Milgrom for remarks.

\end{document}